\documentclass[aps,prb,superscriptaddress,reprint,amsmath,amssymb]{revtex4-1}
\usepackage{amsmath}
\usepackage{amsfonts}
\usepackage{amssymb}
\usepackage{graphicx}
\usepackage{array}

\begin{document}

\title{Onset of Quantum Criticality in the Topological-to-Nematic Transition  \\
in a Two-dimensional Electron Gas at Filling Factor $\nu=5/2$}

\author{K.A. Schreiber}
\affiliation{Department of Physics and Astronomy, Purdue University, West Lafayette, IN 47907, USA}
\author{N. Samkharadze}
\email[]{Present address: QuTech and Kavli Institute of NanoScience, Delft University of Technology, 2600 GA Delft, The Netherlands.}
\affiliation{Department of Physics and Astronomy, Purdue University, West Lafayette, IN 47907, USA}
\author{G.C. Gardner}
\affiliation{School of Materials Engineering, Purdue University, West Lafayette, IN 47907, USA}
\affiliation{Birck Nanotechnology Center Purdue University, West Lafayette, IN 47907, USA}
\author{Rudro R. Biswas}
\affiliation{Department of Physics and Astronomy, Purdue University, West Lafayette, IN 47907, USA}
\author{M.J. Manfra}
\affiliation{Department of Physics and Astronomy, Purdue University, West Lafayette, IN 47907, USA}
\affiliation{School of Materials Engineering, Purdue University, West Lafayette, IN 47907, USA}
\affiliation{Birck Nanotechnology Center Purdue University, West Lafayette, IN 47907, USA}
\affiliation{School of Electrical and Computer Engineering, Purdue University, West Lafayette, IN 47907, USA }
\author{G.A. Cs\'{a}thy}
\email[]{gcsathy@purdue.edu}
\affiliation{Department of Physics and Astronomy, Purdue University, West Lafayette, IN 47907, USA}
\affiliation{Birck Nanotechnology Center Purdue University, West Lafayette, IN 47907, USA}

\date{\today}

\begin{abstract}

Under hydrostatic pressure, the ground state of a two-dimensional electron gas at $\nu=5/2$
changes from a fractional quantum Hall state to the stripe phase. By measuring  
the energy gap of the fractional quantum Hall state and of the onset temperature of the stripe phase
we mapped out a phase diagram of these competing phases in the pressure-temperature plane. 
Our data highlight the dichotomy of two descriptions of the half-filled Landau level near the quantum critical point: 
one based on electrons and another on composite fermions.

\end{abstract}

\maketitle


The fractional quantum Hall state  (FQHS) at the Landau level filling factor $\nu=5/2$
remains one of the most enigmatic ground states of the two-dimensional electron gas
subjected to a perpendicular magnetic field
\cite{firstfivehalf,fivehalf}. This FQHS, similarly to all other FQHSs \cite{tsui,wen}, is
topologically ordered, it is believed to belong to the Pfaffian universality class
\cite{MooreRead,sim1,sim2,sim3,sim4,sim5,sim6,sim7,sim8,anti1,anti2,anti3,anti4},
and it is thought to support non-Abelian excitations \cite{MooreRead}. 
Within the framework of the composite fermion theory \cite{Jain,halperin},
the $\nu=5/2$ FQHS can be understood as being due to pairing
of composite fermions, a pairing driven by the residual attractive interactions  
between the composite fermions
\cite{MooreRead,ReadGreen,greiter,scarola,rezayi,kwon,Sid}. 

Our recent measurements \cite{Samkharadze} revealed that the ground state at $\nu=5/2$ 
undergoes a pressure-driven quantum phase transition from the FQHS to a stripe phase 
\cite{StripeTheory1,StripeTheory2,StripeTheory3,Stripe1,Stripe2,Stripe3,Stripe4,Stripe5,Stripe6,Stripe7,Stripe8}. 
In this experiment the two-dimensional electron gas was not exposed to an in-plane
magnetic field or any other externally applied symmetry breaking fields \cite{Samkharadze}.
This phase transition is intriguing since it does not belong the class of
topological phase transitions in which both phases involved are topologically non-trivial
\cite{class1,class2,class3,tr1,tr2,tr3}.
Indeed, these two phases have fundamentally different orders: the $\nu=5/2$ FQHS 
is topologically ordered, while the stripe phase is a traditional broken symmetry phase supporting nematic order 
\cite{frad,kim17,inti17}.


In this Rapid Communication we present finite temperature measurements of the FQHS and the pressure-induced stripe phase
at $\nu=5/2$. The obtained data allows us to extract energy scales of these two phases,
the energy gap of the $\nu=5/2$ FQHS and the onset temperature of the stripe phase, in particular.
Using these quantities we trace a phase diagram in the pressure-temperature plane.
This phase diagram is a first example of a diagram exhibiting quantum criticality due to the competition 
of a topological and a traditional broken symmetry phase.
Theoretical details of the transition studied have not yet been worked out.
We therefore expect our results to motivate future work on competing phases in topological materials.

\begin{figure} [b]
\includegraphics[width=\columnwidth]{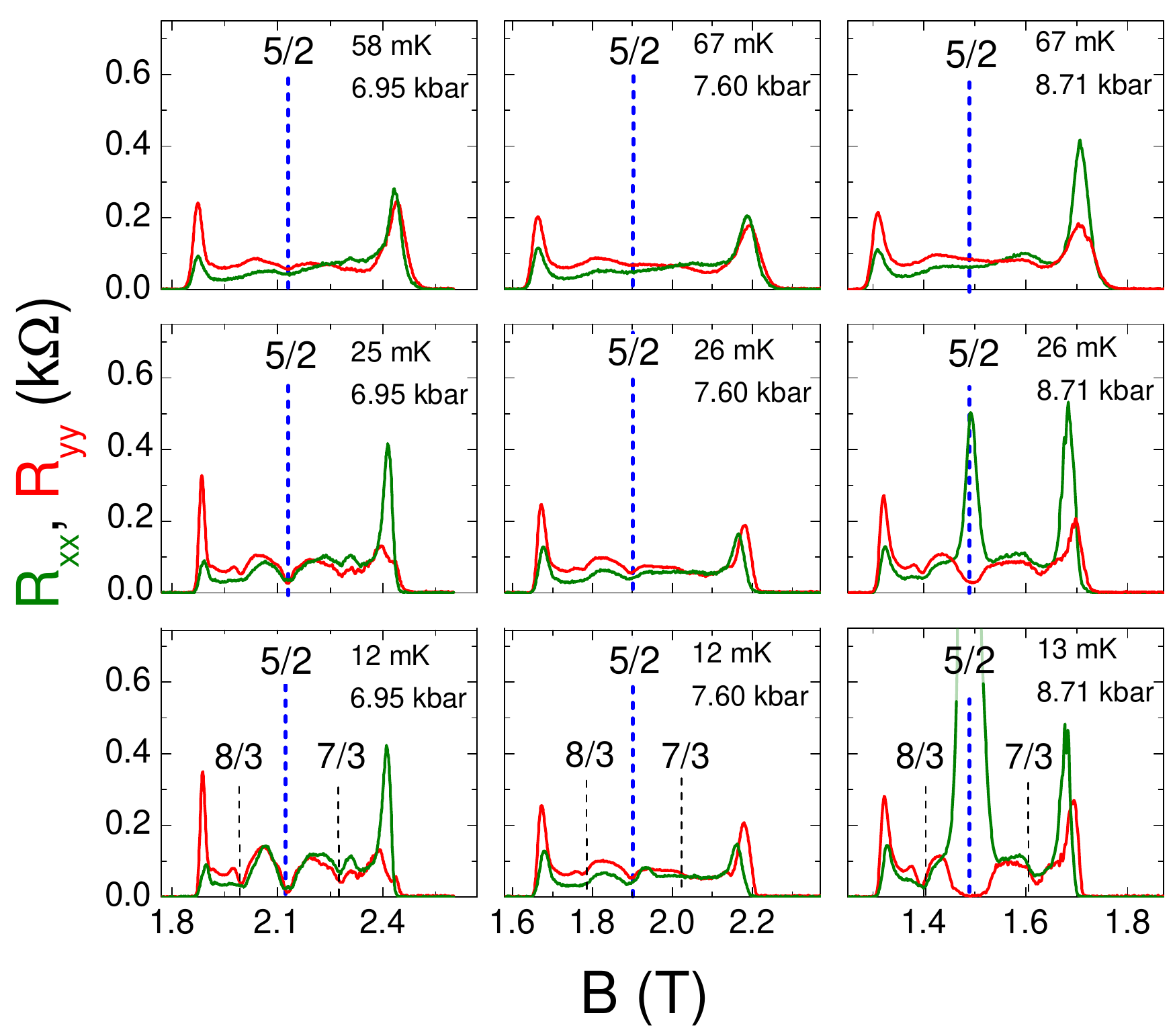}
\caption{The evolution of magnetotransport between $\nu=2$ and $3$ at three temperature and three pressure values. 
The green lines show $R_{xx}$ measured along the [1$\bar{1}$0] crystallographic direction of the GaAs host, while
the red lines $R_{yy}$ measured along [110]. The longer vertical dashed lines mark $\nu=5/2$, while the shorter dotted lines
are at $\nu=7/3$ and $8/3$. The ground state at $\nu=5/2$ and at $6.95$~kbar is a FQHS, at $7.60$~kbar is a nearly isotropic Fermi fluid, and at $8.71$ an electronic stripe phase. Data sets at the lowest tempeature for 6.95 and 7.60~kbar are from Ref.\cite{Samkharadze}.
}
\label{fig1}
\end{figure}

We measured a two-dimensional electron gas with a density
of $n=2.8 \times 10^{11}$~cm$^{-2}$ and mobility of $\mu= 15 \times 10^6$~cm$^2$/Vs
confined to GaAs/AlGaAs quantum well
\cite{Samkharadze,Deng,manf}. The sample was mounted in a pressure clamp cell \cite{pcell}.  
As shown in the Supplemental Material at \cite{suppl}, both the electron density and the mobility decrease with
an increasing pressure \cite{book}. Since the pressure cannot be changed in-situ, the sample is warmed up to room
temperature every time we change the pressure. Pressures quoted throughout this Rapid Communication 
are estimated at low temperatures and temperatures reported are measured on the mixing chamber.

In Fig.1 we show the dependence of the longitudinal magnetoresistance for the Landau filling factor range $2< \nu < 3$
on the temperature and pressure. We have measured the magnetoresistance along two different crystallographic directions
of the GaAs host: $R_{xx}$ is measured along the [1$\bar{1}$0] crystallographic direction, while
$R_{yy}$ along the [110] direction.
Our analysis is focused at $\nu=5/2$, marked by vertical dashed lines in Fig.1. 
In the following  we discuss the three different ground states stabilized at $\nu=5/2$ at different
values of the pressure.

At $P=6.95$~kbar and $T=12$~mK, the longitudinal magnetoresistance near $\nu=5/2$ is vanishingly small and
nearly isotropic. Such a behavior, together with a quantized Hall resistance \cite{Samkharadze},
indicates the presence of a FQHS at $\nu=5/2$ \cite{firstfivehalf,fivehalf}. The density of states of the FQHS at $\nu=5/2$, 
similarly to that of any other FQHS, has an energy gap, hence this FQHS is an incompressible 
quantum liquid \cite{firstfivehalf,fivehalf}.

In contrast, at $P=8.71$~kbar and  $T = 13$~mK,  the longitudinal magnetoresistance at $\nu=5/2$ is strongly anisotropic.
The anisotropic magnetoresistance we observe at $\nu=5/2$ and $P=8.71$~kbar are identical in all aspects
to that of the prototype stripe phase forming at $\nu=9/2$  and at other filling factors
\cite{StripeTheory1,StripeTheory2,StripeTheory3,Stripe1,Stripe2,Stripe3,Stripe4,Stripe5,Stripe6,Stripe7,Stripe8}. 
Indeed, anisotropy at both of these filling factors develops
in the absence of the application of any in-plane $B$-field and
 in a very limited range of filling factors of width 
$\Delta \nu \approx 0.15$ around the half-integer value
\cite{Stripe1,Stripe2,Stripe3,Stripe4,Stripe5,Stripe6,Stripe7,Stripe8,Samkharadze}. 
In the absence of an energy gap in the density of states, the stripe phase at $\nu=5/2$  
is compressible \cite{Stripe1,Stripe2,Stripe3,Stripe4,Stripe5,Stripe6,Stripe7,Stripe8}.

The stripe phase we observe at $\nu=5/2$ is likely related to nematic phases
developing in the two-dimensional electron gas but under different experimental conditions,
specifically with the application of an in-plane $B$-field.
Indeed, nematicity at $\nu=5/2$ is induced in the presence of a sizable in-plane $B$-field but,
in contrast to our observations, 
anisotropy under tilt develops over a very wide $\Delta \nu \approx 0.6$ range \cite{tilt1,tilt2,tilt3,tilt4,tilt5}.
Nematic phases are also reported in Refs.\cite{tilt6,tilt7} but these phases, in contrast to
the stripe phase we observe, are incompressible.

At $P=7.60$~kbar and  $T = 12$~mK, the magnetoresistance at $\nu=5/2$ remains finite, 
featureless, and nearly isotropic \cite{Samkharadze}. Such a behavior signals that
the ground state cannot be a FQHS nor a stripe phase. 
A similar behavior of the magnetoresistance was observed at $\nu=1/2$ and was associated with Fermi liquid 
behavior \cite{RuiDu}. We therefore interpret our data at $P=7.60$~kbar and  $T = 12$~mK as
evidence for a Fermi liquid-like state. 

Magnetoresistance data shown in the lowest row of panels of Fig.1 demonstrates that
the ground state at $\nu=5/2$ as measured near $12$~mK evolves
from a FQHS toward an electronic stripe phase as the pressure is increased \cite{Samkharadze}. 
Fig.1 also shows how a rising temperature changes the magnetoresistance at $\nu=5/2$. 
As a rule, at a higher temperature features of the magnetoresistance become less pronounced. 
For example, at $P=6.95$~kbar there is an increase of the
magnetoresistance at $\nu=5/2$ as the temperature is raised from 12 to 25~mK.
This indicates an enhanced generation of thermally activated excitations in the FQHS.
In addition, at $P=8.71$~kbar the degree of anisotropy of the stripe phase measured 
at $T=26$~mK is weaker than that measured at $T=13$~mK. 

In order to describe the temperature evolution of the observed ground states, we extract a 
characteristic energy scale associated with them.
A FQHS is characterized by the energy gap $\Delta$ of the excitations with respect to the ground state. 
The longitudinal magnetoresistance  in the presence of an energy gap $\Delta$ in the density of states
is proportional to $ \exp (-\Delta/2 k_B T)$, a relationship 
often referred to as the activated behavior. Fig.2a shows the activated behavior of the FQHS at $\nu=5/2$
and also the extracted energy gaps $\Delta$ of the $\nu=5/2$ FQHS at $P=2.58$ and $6.95$~kbar.
We find that the energy gap of the $\nu=5/2$ FQHS decreases with an increasing pressure.

\begin{figure}[t]
\includegraphics[width=\columnwidth]{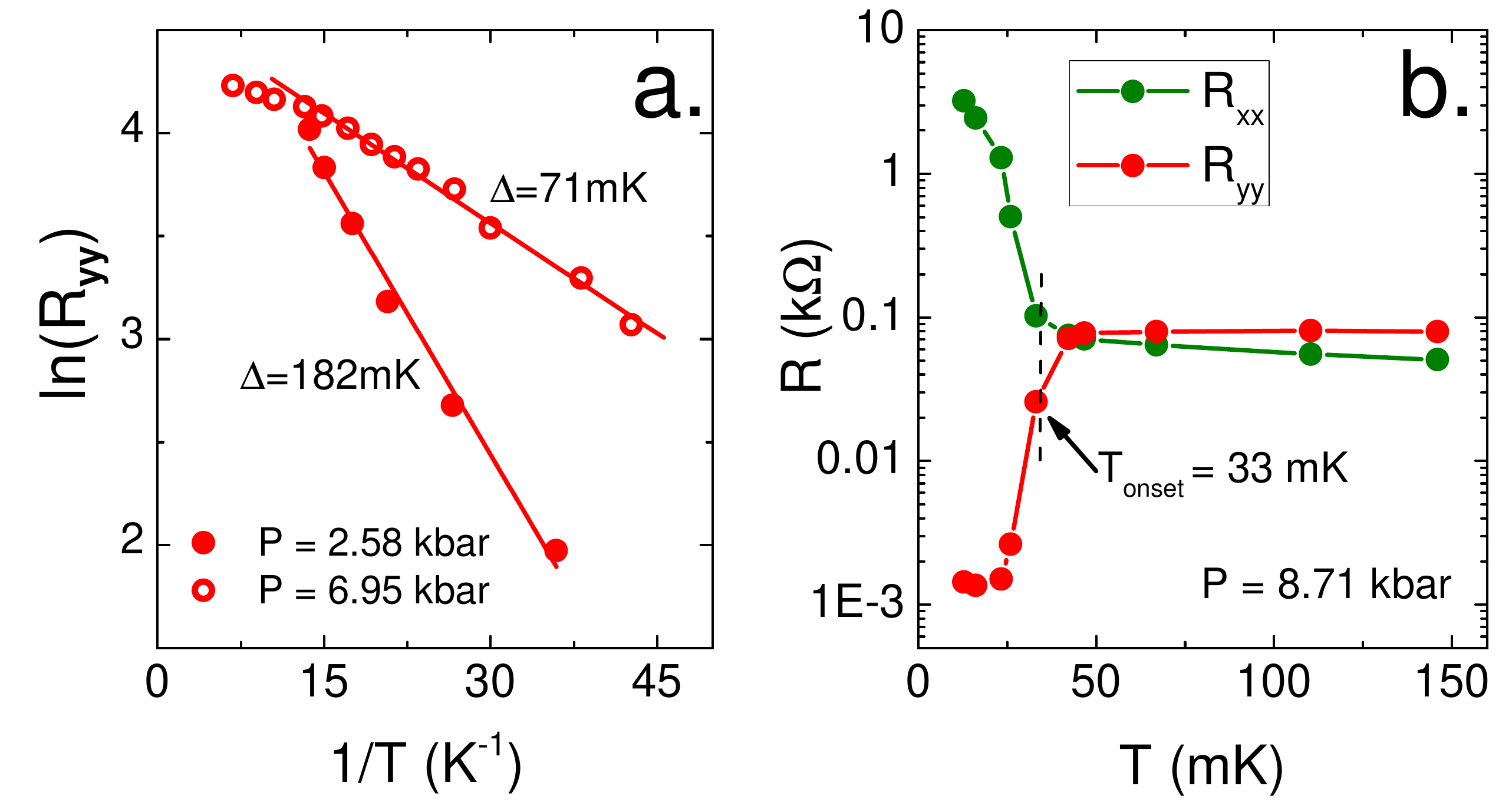}
\caption{Panel a: Arrhenius plots of the $T$-dependent magnetoresistance at $\nu=5/2$ used to extract the energy gap 
$\Delta$ of the $\nu=5/2$ FQHSs.
Panel b: The temperature dependence of the magnetoresistance of the
stripe phase at $\nu=5/2$ shows the development of a large anisotropy.
The vertical dashed line marks the onset temperature for nematicity  $T_{\text{onset}}$.
}
\label{fig2}
\end{figure}

The temperature dependence of the stripe phase at $\nu=5/2$
is shown for $P=8.71$~kbar in Fig.2b. At relatively high temperatures, exceeding
40~mK, the magnetoresistance is nearly isotropic. In our sample we observe a small difference between 
$R_{xx}$ and $R_{yy}$ which is
often seen in experiments and is commonly attributed to imperfections in the sample geometry.
Indeed, since the side of our sample is only 2~mm long and  the Indium ohmic contacts are applied 
by soldering, there is likely a small geometric difference between the $xx$ and $yy$ sides of the sample.
While we do not observe any obvious signatures of density gradients in our sample \cite{pan05},
it is possible that small variations around the mean pressure result in small density fluctuations
which may also influence the magnetoresistance.
In contrast to the behavior of $R_{xx}$ and $R_{yy}$ taken above 40~mK,
$R_{xx}$ and $R_{yy}$ sharply deviate one from another at lower temperatures \cite{Stripe1,Stripe2,Stripe3,Stripe4}. 
As seen in Fig.2b, the $R_{xx}/R_{yy}$ ratio of the resistances in the
two different crystallographic directions exceeds three orders of magnitude at the lowest temperatures.
The relatively abrupt onset of anisotropy is a hallmark property for the stripe phase and
it defines the onset temperature for nematicity $T_{\text{onset}}$. 
We estimate $T_{\text{onset}}$ by imposing a significant anisotropy $R_{xx}=2R_{yy}$ in the linearly interpolated data. 
The dashed line in Fig.2b marks $T_{\text{onset}}$ obtained this way at $P=8.71$~kbar.

\begin{figure}[t]
\includegraphics[width=\columnwidth]{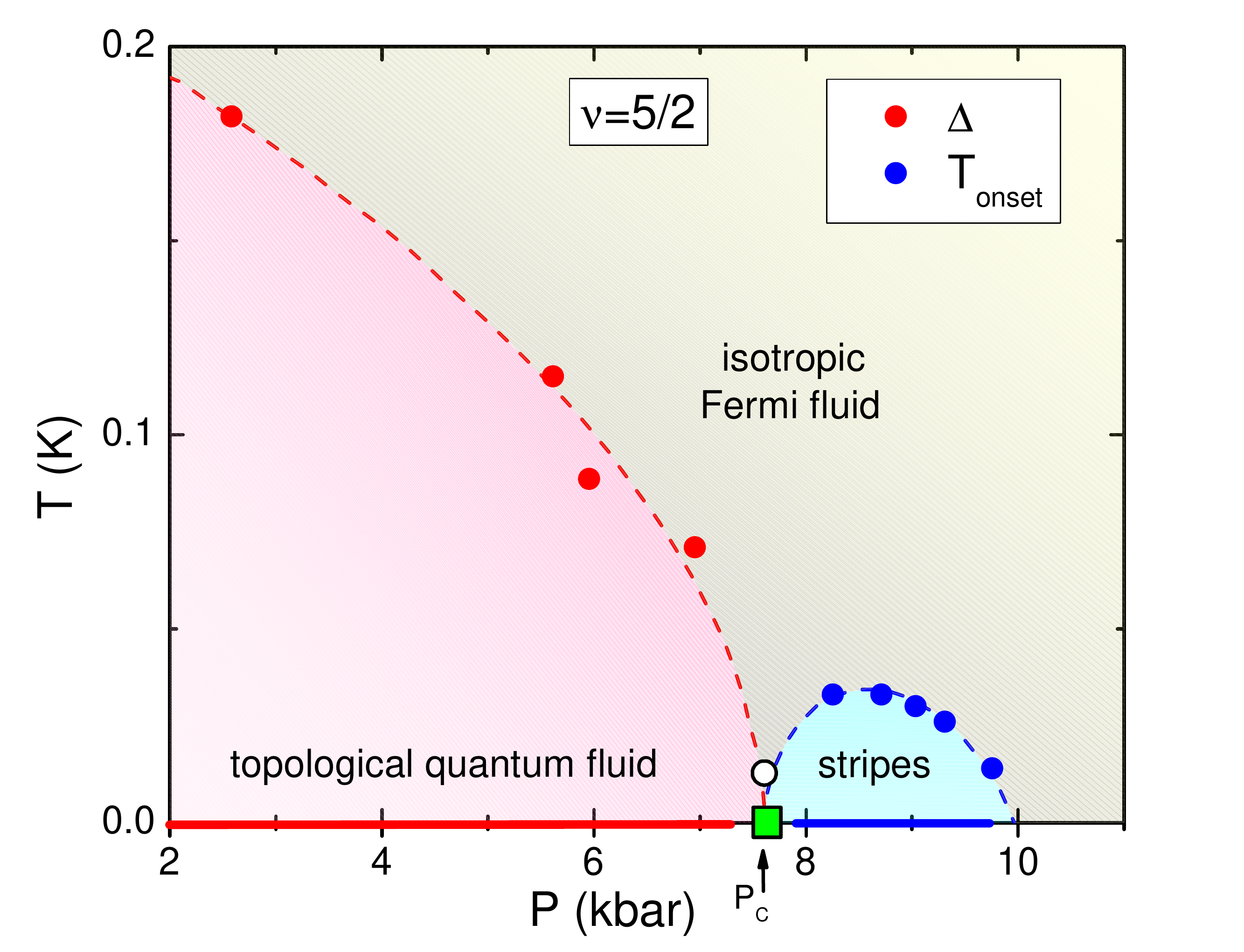}
\caption{A diagram summarizing the behavior at $\nu=5/2$ in the $P$-$T$ phase space. Full symbols represent the
energy gap of the FQHS (red symbols) and the onset temperature of the stripe phase (blue symbols). The open symbol
at $P=7.60$~kbar and $T=12$~mK shows that at these parameters we observe a nearly-isotropic Fermi fluid. 
Dashed lines are guides to the eye. The green square is a quantum critical point.
}
\label{fig3}
\end{figure}

The dependence on the pressure of the energy gap of the $\nu=5/2$ FQHS and of the estimated onset temperature
of the stripe phase at $\nu=5/2$ are summarized in Fig.3. We observe that the energy gap of the $\nu=5/2$ FQHS
is monotonically suppressed with an increasing pressure. At higher pressures we find that the stripe phase is stabilized at $\nu=5/2$. In Fig.3 the dashed red line is a guide to the eye for the energy gap of the $\nu=5/2$ FQHS  
and the dashed blue line for the onset temperature of the stripe phase at $\nu=5/2$. 

Fig.3 can be understood as a phase diagram. Far below the dashed lines the ground state is either the
FQHS or the stripe phase. Above the dashed lines there is the Fermi liquid-like phase.
We note that the red dashed line is not a sharp phase boundary, 
but it represents a crossover between the FQHS and the Fermi liquid.
The blue dashed line denotes a transition of an unknown type.
The continuous horizontal red line at $T=0$ indicates
the ground state is the $\nu=5/2$ FQHS, while the continuous blue line represents the stripe phase
in the limit of $T=0$. Above the dashed lines we have an isotropic Fermi liquid-like phase.
Since data sets at $P=7.60$~kbar are consistent with a Fermi liquid-like state,
the Fermi liquid is wedged in between the FQHS and the stripe, down to at least 12~mK.
The open circle at $P=7.60$~kbar at $T=12$~mK in Fig.3 marks this point of lowest temperature Fermi liquid 
we accessed. Because the Fermi liquid is wedged in between the two ordered phases,
the stripe region forms a dome in the $P$-$T$ phase diagram.

The phase diagram shown in Fig.3 is a first example of an experimentally
obtained diagram exhibiting quantum criticality
of competing topological and a nematic orders.  
While the stripe phase is a well known broken symmetry phase, 
the $\nu=5/2$ FQHS is beyond Landau's description. One cannot associate a local order parameter with this state; 
it is said that a FQHS is a topological quantum liquid \cite{wen}. In fact the topological quantum liquid at
$\nu=5/2$ is different from other FQHSs as it is thought to belong to the Pfaffian universality class
 \cite{MooreRead,sim1,sim2,sim3,sim4,sim5,sim6,sim7,sim8,anti1,anti2,anti3,anti4}.
This phase diagram in the vicinity of $P=7.6$~kbar is very similar to the diagram of a quantum phase transition \cite{sachdev}.
Earlier we suggested a direct quantum phase transition between these two phases  which occurs
at the quantum critical point $P_{C}=7.8 \pm 0.2$~kbar \cite{Samkharadze}. 
This critical point is of a novel type because one of the phases is topological in nature.
As the quantum critical point is crossed with an increasing pressure, the topological order of the FQHS
is destroyed while the nematic order is acquired. 

Obtaining more detailed data near $P_C$ is quite challenging due to the inability to change the
pressure in-situ, a limitation of the technique we use. We emphasize that 
a direct phase transition at $T=0$ remains the simplest, most elegant interpretation of our data.
Any cut in the phase diagram at a finite temperature below the onset of stripes will reveal
the FQHS, Fermi liquid, and stripe sequence of phases as the pressure is increased. 

We think that the phase competition shown in Fig.3 originates from a delicate tuning of
the effective electron-electron interaction with pressure.
Early numerical work provided the insight that a transition from a FQHS to a stripe phase 
is possible at half filling when
the interaction between the electrons deviates from its Coulomb expression \cite{rezayi,wan,wang}.
However, until very recently \cite{Samkharadze} it was not clear whether such peculiar interactions favoring the stripe phase 
can be experimentally realized at $\nu=5/2$. One way to tune the effective electron-electron interaction
is through changing the effective width of the quantum well \cite{rezayi}.  It was later
found that besides the quantum well width, the Landau level mixing parameter \cite{yoshi} 
must also be constrained to stabilize the stripe phase at $\nu=5/2$  \cite{Samkharadze}.
We think one of these effects or perhaps their combination
is responsible for the stabilization of the stripe phase.

A model for the transition from the FQHS to a nematic phase at half filling has recently been formulated \cite{frad}.
In this model, the nematic phase is stabilized by a quadrupolar interaction between the electrons.
In the presence of this type of interaction, it is found that the Fermi liquid behavior can be destroyed either by
fluctuations in the Chern-Simons gauge fields or by the nematic order parameter \cite{frad}. 
As a result, a direct quantum phase transition from the paired FQHS to the nematic phase was obtained \cite{frad}.
Transitions from a FQHS to stripes at half-filled Landau level were also found in theoretical work considering the
interplay between nematic and gauge fluctuations \cite{kim17}
and when a changing mass anisotropy is present in the system \cite{inti17}.

Nonetheless, our understanding of the phase transition near the quantum critical point is still incomplete.
Below the critical pressure, a FQHS requires the existence of composite fermions \cite{MooreRead}. 
In contrast, composite fermions are not required to account for the stripe phase above the critical pressure
\cite{StripeTheory1,StripeTheory2,StripeTheory3}. 
The existence of a quantum critical point in Fig.3 thus highlights the dichotomy of the two
descriptions of the half-filled Landau level: one based on electrons \cite{StripeTheory1,StripeTheory2,StripeTheory3}
and another on composite fermions \cite{Jain,halperin}.

Interest in the half-filled Landau level was recently rekindled by theories according to which
the composite fermions are Dirac-like at exact particle-hole symmetry \cite{dirac1,dirac2,dirac3,dirac4}. 
These theories naturally account for a Fermi sea and for a FQHS at half-filling, 
but do not accomodate the formation of the stripe phase \cite{dirac1}.
In our experiment particle-hole symmetry is broken due to significant Landau level mixing and finite width effects,
therefore these theories most likely do not strictly apply.

Finally, we note that the FQHSs developing at $\nu=7/3$ and $8/3$ deteriorate
near the quantum critical point.
Indeed, data from Fig.1 at $P=6.95$~kbar and $8.71$~kbar the presence of depressions in the magnetoresistance
at $\nu=7/3$ and $8/3$ at the lowest temperatures reached indicates weak FQHSs at these filling factors. 
However, at the intermediate pressure $P=7.60$~kbar and $T=12$~mK these weak depressions at $\nu=7/3$
and $8/3$ have virtually disappeared. In the vicinity of the critical pressure we thus observe a conspicuous
loss of electronic correlations responsible for the $\nu=7/3$ and $8/3$ FQHSs. One possibility is that such a
deterioration of the FQHSs at $\nu=7/3$ and $8/3$ near the quantum critical point 
could be due to enhanced quantum fluctuations.  

To conclude, we have measured the pressure-dependent energy gap of the FQHS at $\nu=5/2$
and the onset temperature of the stripe phase developing at the same filling factor.
These quantities allowed us to map out the phase diagram 
near the instability of the parent Fermi sea toward a FQHS and toward stripes in the $P$-$T$ parameter space.
We found that finite temperature measurements corroborate with the interpretation
of a direct phase transition from the FQHS to the stripe phase in the limit of zero temperatures.
We have thus demonstrated that the two-dimensional electron gas at $\nu=5/2$ is a
model system which supports competing topological and traditional nematic orders
in the $P$-$T$ parameter space.

This work was supported by the US Department of Energy, Office of Basic Energy Sciences, Division of Materials 
Sciences and Engineering under the award DE-SC0006671.


\end{document}